\documentclass[twocolumn]{revtex4}
\usepackage{graphicx,multirow,array,amsmath,empheq,amsthm,amssymb,amsfonts,tikz,ragged2e,color}
\begin{document}
\title{\textcolor{violet}{Shear Alfv\'enic Waves in Electron-Positron Plasma: Analysis of Solitary Wave Properties and Numerical Validation}}
%\title{\textcolor{violet}{Shear Alfv\'enic Waves in a Magnetized Pair Plasma Medium: Analysis of Solitary Wave Properties and Numerical Validation}}
%"Nonlinear Propagation of Shear Alfvénic Waves in Electron-Positron Plasma: Analysis of Solitary Wave Properties and Numerical Validation"
\author{T. I. Rajib}
\date{\today}
\address{Department of Physics, Jahangirnagar University, Savar, Dhaka-1342, Bangladesh\\
%\address{Institute for Quantum Science and Engineering, Department of Physics and Astronomy, Texas A$\&$M University, College Station-77843, Texas, USA\\
%*Department of Physics, Jahangirnagar University, Savar, Dhaka-1342, Bangladesh\\
E-mail Address: tirajibphys@gmail.com OR tirajibphys@juniv.edu}

%\begin{sloppypar}
\begin{abstract}
%The reductive perturbation approach was used to explore the nonlinear propagation of shear  Alfv\'enic waves in an EP plasma medium. The solitary wave solution of the derivative nonlinear Schr\"{o}dinger %equation is used to identify the basic properties of EP shear Alfv\'{e}n (EPSA) waves. The basic features (i.e., speed, amplitude, and width) of the EPSA waves are analyzed. It is also examined that hump %shape solitary waves are found. Moreover, the numerical result obtained by the finite difference method is compared with the exact solution. It is seen that this is in good agreement with each other. The %significance of our findings is in understanding the nonlinear electromagnetic wave phenomena in laboratory plasma and space environments where EP plasma exists.
The investigation employed the reductive perturbation approach to study the nonlinear propagation of shear Alfvénic waves in an electron-positron (EP) plasma medium. Utilizing the solitary wave solution of the derivative nonlinear Schrödinger equation, the fundamental properties of EP shear Alfvén (EPSA) waves were identified, with a focus on their key features such as speed, amplitude, and width. The analysis revealed the presence of hump-shaped solitary waves. Additionally, comparing the numerical results obtained through the finite difference method and the exact solution demonstrated good agreement between the two. These findings hold significance in comprehending nonlinear electromagnetic wave phenomena in laboratory plasma and space environments where EP plasma is present such as solar wind, Earth's magnetosphere, pulsars' magnetospheres, microquasars, and tokamaks.
\end{abstract}

\maketitle

\section{Introduction}
%The electron-positron (EP) plasmas are observed in
%various astrophysical environments like the solar wind \cite{Clem2000,Adriani2009,Adriani2011}, the magnetosphere of Earth \cite{Ackermann2012,Aguilar2014}, pulsars magnetosphere \cite{Profumo2012}, %and microquasars \cite{Siegert2016}. In addition, this pair of plasmas can be
%produced by imposing an ultra-intense laser on the matter in
%the laboratory \cite{Sarri2013,Sarri2015}. It is also mentioned that EP plasmas are created in the presence
%of strong magnetic/electric fields or extremely high
%temperatures \cite{Thoma2009}. The progress of high-efficiency
%techniques for assembling pure positron plasmas in Penning traps
%\cite{Greaves1994,Surko1989} now make laboratory experiments
%able to produce this type of pair plasmas. It is shown that EP pair
%production is expected to occur in post-disruption plasmas in
%large tokamaks \cite{Helander2003}. The physics of an EP
%plasma is quite different from electron-ion plasmas due to the
%large ion-to-electron mass \cite{Swanson2003,Stix1992}. The EP
%plasmas at the surface of magnetars and fast-rotating neutron stars are held in strong magnetic fields while superseding
%magnetic fields can be created in intense laser-plasma
%interaction experiments. Therefore, the understanding of
%collective phenomena in magnetized pair plasmas has been a topic
%of significant interest \cite{Zank1995,Iwamoto1993,Lominadze1982,Gedalin1985,Yu1986,Brodin2007,Rajib2015}.\\

Electron-positron (EP) plasmas are observed in diverse astrophysical contexts, including the solar wind \cite{Clem2000,Adriani2009,Adriani2011}, Earth's magnetosphere \cite{Ackermann2012,Aguilar2014}, pulsars' magnetospheres \cite{Profumo2012}, and microquasars \cite{Siegert2016}. In laboratory settings, EP plasmas can be generated through ultra-intense laser-matter interactions \cite{Sarri2013,Sarri2015}. Strong magnetic/electric fields or extremely high temperatures also contribute to EP plasma creation \cite{Thoma2009}, with advancements in assembling pure positron plasmas in Penning traps \cite{Greaves1994,Surko1989}. The occurrence of EP pair production is anticipated in post-disruption plasmas in large tokamaks \cite{Helander2003}. Notably, the physics of EP plasmas differs significantly from electron-ion plasmas due to the large ion-to-electron mass ratio \cite{Swanson2003,Stix1992}. EP plasmas in the vicinity of magnetars and fast-rotating neutron stars experience strong magnetic fields, with similar conditions replicable in intense laser-plasma interaction experiments. Consequently, the study of collective phenomena in magnetized pair plasmas has garnered substantial interest \cite{Zank1995,Iwamoto1993,Lominadze1982,Gedalin1985,Yu1986,Brodin2007,Rajib2015,Rajib2022a,Rajib2022b}.

Extensive theoretical and numerical investigations have elucidated the distinctive physics of EP plasmas, covering fundamental wave physics \cite{Zank1995}, reconnection \cite{Bessho2005,Blackman1994,Yin2008}, and nonlinear solitary waves \cite{Berezhiani1994,Cattaert2005}. Sakai and Kawata \cite{Sakai1980} analyzed small-amplitude solitary waves using higher-order modified Korteweg-de Vries (mK-dV) equations from relativistic hydrodynamic equations, while Zank and Greaves \cite{Zank1995} explored linear properties of electrostatic and electromagnetic modes in both unmagnetized and magnetized pair plasmas. El-Wakil \textit{et al.} \cite{Elwakil2019} investigated solitary wave propagation in EP pair plasmas using reductive perturbation theory and nonlinear equations. Iwamoto \cite{Iwamoto1993} provided a kinetic description of various linear collective modes in a non-relativistic pair magnetoplasma, and Helander and Ward \cite{Helander2003} demonstrated positron creation in tokamaks due to collisions of runaway electrons with plasma ions/thermal electrons. Lontano \textit{et al.} \cite{Lontano2001} studied the interaction between arbitrary amplitude electromagnetic fields and EP (hot) plasma. The nonlinear wave propagation of EP plasma in a pulsar magnetosphere was investigated by Kennel \textit{et al.} \cite{Kennel1976,Arons1986}, who observed that large amplitude waves determine average plasma properties.

Shear Alfvén (SA) waves are magnetohydrodynamic waves in magnetized plasmas that propagate in the direction of the ambient magnetic field, and the motion of the plasma particles (here EP) and the perturbation of the magnetic field are transverse to the direction of propagation. Observations of SA waves have been documented in numerous laboratory experiments \cite{Gekelman1999} and occur naturally in a diverse range of astrophysical magnetized plasmas, including planetary magnetospheres \cite{Pater2001}, Earth's aurora \cite{Louarn1994}, solar corona \cite{Hollweg1982}, and more. Due to their propagation primarily along an ambient magnetic field \cite{Hasegawa1976,Goertz1979,Gekelman1997}, SA wave demonstrates efficient energy transport. Various theoretical studies \cite{Sakai1980,Melikidze1981,Stenflo1985} have explored nonlinear Alfvén wave propagation in EP plasmas. Melikidze \textit{et al.} \cite{Melikidze1981} analyzed solitary wave properties for field-aligned electromagnetic waves, while Stenflo \textit{et al.} \cite{Stenflo1985} considered Alfvén waves in a cold electron-positron plasma, describing the coupling of radiation with cold electrostatic oscillations. Mikhailovskii \textit{et al.} \cite{Mikhailovskii1985} theoretically examined nonlinear Alfvén waves in a relativistic EP plasma.

%%%%%%%%%%%%%%%%%%%%%%%%%%%%%%%%%%%%%%%%%%%%%%%%%%%%%%%%%%%%%%%%%
%Apart from the opposite polarity EP pair plasma, recently, based on the theoretical predictions and
%satellite/experimental observations, several authors
%\cite{D'Angelo2001,D'Angelo2002,Shukla2006,Mamun2002,Sayed2007,Rahman2008,Mamun2008a,Mamun2008b,Verheest2009}
%have considered a dusty plasma with dust of opposite polarity, and
%have investigated linear
%\cite{D'Angelo2001,D'Angelo2002,Shukla2006} and nonlinear
%\cite{Mamun2002,Sayed2007,Rahman2008,Mamun2008a,Mamun2008b,Verheest2009}
%electrostatic waves excluding
%\cite{D'Angelo2001,Shukla2006,Mamun2002,Sayed2007,Rahman2008,Mamun2008a,Mamun2008b,Verheest2009}
%or including \cite{D'Angelo2002} external static magnetic field. On
%the other hand, Shukla \cite{Shukla2004} and Mamun
%\cite{Mamun2011} have considered a medium of magnetized opposite
%polarity dust to study linear and nonlinear
%electromagnetic waves. Shukla has analytically studied linear
%dispersive dust Alfv\'en waves, and associated dipolar vortex
%\cite{Shukla2004}, whereas Mamun has examined the nonlinear
%propagation of the fast and slow dust magnetoacoustic (MA) modes
%\cite{Mamun2011}.

In addition to opposite polarity EP pair plasmas, recent theoretical predictions and observational data from satellites and experiments have prompted several authors \cite{D'Angelo2001,D'Angelo2002,Shukla2006,Mamun2002,Sayed2007,Rahman2008,Mamun2008a,Mamun2008b,Verheest2009} to explore dusty plasmas featuring dust particles with opposite polarity. These studies have delved into both linear \cite{D'Angelo2001,D'Angelo2002,Shukla2006} and nonlinear \cite{Mamun2002,Sayed2007,Rahman2008,Mamun2008a,Mamun2008b,Verheest2009} electrostatic waves, with investigations conducted both in the absence \cite{D'Angelo2001,Shukla2006,Mamun2002,Sayed2007,Rahman2008,Mamun2008a,Mamun2008b,Verheest2009} and presence \cite{D'Angelo2002} of an external static magnetic field. Moreover, Shukla \cite{Shukla2004} and Mamun \cite{Mamun2011} have explored a medium composed of magnetized dust particles with opposite polarity to investigate both linear and nonlinear electromagnetic waves. Shukla's analytical work focused on linear dispersive dust Alfvén waves and associated dipolar vortices \cite{Shukla2004}, while Mamun's research delved into the nonlinear propagation of fast and slow dust magnetoacoustic modes \cite{Mamun2011}.

%Shear Afv\'en waves (SAW) have been observed in several laboratory experiments\cite{} and nature in a wide variety of astrophysical magnetized plasmas such
%as planetary magnetospheres\cite{}, Earth's aurora\cite{}, solar corona\cite{}, etc. Because of its propagation nearly along an ambient magnetic field \cite{}, SAW can transport energy
%efficiently.

%In the earlier work \cite{Mamun2013}, Mamun studied the shear  Alfv\'enic solitary structures in an opposite
%polarity dusty plasma medium. Mamun concluded that this plasma model is applied to any opposite polarity plasma medium such as EP and Electron-ion. This research article deals %with the extension of the earlier one \cite{Mamun2013} with the ratio of positron mass to electron mass is equal to unity i.e., $\alpha=m_p/m_e=1$. We, in our present work, consider %a medium of EP plasma and make a systematic analysis of high-frequency nonlinear shear Alfv\'en solitary waves by using a well-known DNSE equation with reductive perturbation %technique.

In a prior study \cite{Mamun2013}, Mamun explored shear Alfv\'enic solitary structures within an opposing polarity dusty plasma medium, noting the applicability of this plasma model to various scenarios, including EP and electron-ion plasmas. This current research builds upon the earlier work \cite{Mamun2013}, specifically focusing on the extension of the study when the ratio of positron mass to electron mass is set to unity, denoted by $\alpha=m_p/m_e=1$. In our present investigation, we delve into the realm of EP plasma, conducting a systematic analysis of high-frequency nonlinear shear Alfv\'en solitary waves through the utilization of the well-established derivative nonlinear Schrödinger equation (DNSE) coupled with the reductive perturbation technique.

%The layout of the manuscript is as follows: The governing equations describing our plasma model and a brief description of the mathematical technique, followed by the derivation of a DNSE equation with its %solution are given in Section \ref{GE}. The numerical analysis and results are presented in Section \ref{NA}, and finally, a discussion is drawn in Section \ref{D}.

The manuscript is structured as follows: Section \ref{GE} provides the governing equations outlining our plasma model and a concise explanation of the mathematical technique, followed by the derivation of a DNSE along with its solution. Section \ref{NA} presents the numerical analysis and results, and lastly, Section \ref{D} concludes with a discussion.

\section{Governing Equations}
\label{GE}
We consider collisionless, magnetized ultra-high frequency electromagnetic perturbations in a
medium of EP, which is assumed to be immersed in an external static magnetic field $\vec{B}_0$ (i.e., $\vec{B}_0$$||$$\hat{z}$, where $\hat{z}=$ unit vector along the z-axis). Thus, the macroscopic state of the medium of
opposite polarity magnetized EP fluids
\cite{D'Angelo2002,Shukla2004,Mamun2011} is described by the following set of equations
\begin{eqnarray}
&&\hspace*{5mm}{\partial_t}N_s + {\partial_z} (N_s{ U}_{sz}) = 0,\label{1}\\
&&\hspace*{5mm}m_sD_t^{s}{\vec U}_s =q_{s}\left({\vec E} +\frac{1}{c}{\vec U}_s\times{\vec B}\right),\label{2}\\
&&\hspace*{5mm}{\hat{z}}\times\partial_z {\vec E} = - \frac{1} {c}{\partial_t {\vec B}},\label{3}\\
&&\hspace*{5mm}{\hat{z}}\times\partial_z {\vec B} = \frac{4 \pi}{c}\sum_{s} q_{s}  N_s{\vec U}_s,\label{4}\\
&&\hspace*{5mm}\partial_z{\vec B} = 0, \label{5}\\
&&\hspace*{5mm}0 = 4 \pi \sum_{s} q_{s} N_{s},\label{6}
\end{eqnarray}
where $D_t^{s}=\partial_t +U_{sz}\partial_z$, $\partial_z={\partial }/{\partial_z}$ and  $\partial_t={\partial }/{\partial_t}$
$s~(= e, p)$ denotes the species (namely, electron and
positron); $m_s$, $q_s$, and $N_s$ are, respectively, mass,
charge, and number density of the species $s$;  ${\vec U}_s$ is
the hydrodynamic velocity; ${\vec E}$ is the electromagnetic wave electric field, and
${\vec B}$ is the sum of external and wave magnetic fields; $c~(=3\times10^8~m/s)$ is the speed of light in a
vacuum.

Now, neglecting the contribution of the displacement
current as of the wave phase speed is negligible compared to the
speed of light $c$ and assuming the quasi-neutrality condition
($N_e=N_p$), we can reduce our basic equations
(\ref{1})$-$(\ref{6}) to
\begin{eqnarray}
&&\hspace*{-5mm}{\partial_t}N_s + {\partial_z} (N_s{ U}_{sz}) = 0,\label{7}\\
&&\hspace*{-5mm}N_sD_t^{s}\left[(1+\alpha){U}_{sx} -{{\alpha}{\cal H}({\partial_z}B_y})\right]-b{{\partial_z}B_x}=0, \label{8}\\
&&\hspace*{-5mm}N_sD_t^{s}\left[(1+\alpha){U}_{sy} +{{\alpha}{\cal H}({\partial_z}B_x})\right]-b{{\partial_z}B_y}=0, \label{9}\\
&&\hspace*{-5mm}N_sD_t^{s}(1+\alpha){\vec U}_{sz}+\frac{b}{B_0}(B_y{{\partial_z}B_y}+B_x{{\partial_z}B_x})=0, \label{10}\\
&&\hspace*{-5mm}{\partial_t B_x}-\partial_z(U_{sx}B_0-U_{sz}B_x)+\frac{B_0}{\omega_{cs}}\partial_zD_t^{s}U_{sy}=0,\label{11}\\
&&\hspace*{-5mm}{\partial_t B_y}-\partial_z(U_{sy}B_0-U_{sz}B_y)-\frac{B_0}{\omega_{cs}}\partial_zD_t^{s}U_{sx}=0,\label{12}
\end{eqnarray}
where $B_z=B_0$, $\omega_{cs}=q_sB_0/m_sc$, $V_{As}=B_0/\sqrt{4\pi N_{s0}m_s}$, $b= V_{As}^2N_{s0}/B_0$, ${\cal H}=b/{\omega_{cs}N_s}$, $\alpha=m_p/m_e=1$ (Note that we consider $\alpha$ is equal to unity on the rest of the manuscript for our considered system as the mass of electron and positron is same).

We are interested in high-frequency small but finite amplitude electromagnetic
perturbation modes propagating along the $z$-axis (i.e., all
dependent variables depend on $z$ and $t$ only) in the presence of an
external static magnetic field ${\vec B}_0$ that are exactly parallet 
$\theta$ with the $z$-axis (i.e., $\theta=0^{\circ}$). We construct a weakly nonlinear theory
which leads to the following stretched coordinates \cite{Washimi1966,Haider2019}
\begin{eqnarray} \left.
\begin{array}{l}
\xi=\epsilon(z-V_pt),\\
\tau=\epsilon^{2}t,
\end{array}
\right\} \label{13}
\end{eqnarray}
where $V_p$ is the phase speed  of the EPSA waves and
 $\epsilon$ $(0<\epsilon<1)$ is a small parameter measuring the weakness of the dispersion. We then expand the dependent variables $N_s$, $U_{sx}$, $U_{sy}$, $U_{sz}$, $B_x$, and $B_y$ (where the subscript $x$, $y$, and $z$, respectively, represent
$x$-, $y$- and $z$-components of the quantity involved) about their equilibrium values in powers of
$\epsilon$:
\begin{eqnarray} \left.
\begin{array}{l}
N_s=1+\epsilon N_s^{(1)}+\epsilon^{2}N_s^{(2)}+ \cdot \cdot \cdot,\\
U_{sx}=0+\epsilon^{1/2} U^{(1)}_{sx}+\epsilon^{3/2}U_{sx}^{(2)}+\cdot \cdot \cdot,\\
U_{sy}=0+\epsilon^{1/2} U_{sy}^{(1)}+\epsilon^{3/2}U_{sy}^{(2)}+\cdot \cdot \cdot,\\
U_{sz}=0+\epsilon U^{(1)}_{sz}+\epsilon^{2}U_{sz}^{(2)}+ \cdot \cdot \cdot,\\
B_{x}=0+\epsilon^{1/2} B^{(1)}_{x}+\epsilon^{3/2}B_{x}^{(2)}+\cdot \cdot \cdot,\\
B_{y}=0+\epsilon^{1/2} B^{(1)}_{y}+\epsilon^{3/2}B_{y}^{(2)}+\cdot \cdot \cdot,\\
\end{array}
\right\} \label{14}
\end{eqnarray}

Now, substituting the above-stretched co-ordinates (\ref{13})
and expansion series into  equations (\ref{8}), (\ref{9}), (\ref{11}), and (\ref{12}) and equating the coefficients of
$\epsilon^{3/2}$ we obtain a set of equations
that can be simplified as
\begin{eqnarray}
&&\hspace*{-1mm} U^{(1)}_{sx}=-V_pB_{x}^{(1)}/B_0,\label{15}\\
&&\hspace*{-1mm} U^{(1)}_{sy}=-V_pB_{y}^{(1)}/B_0,\label{16}\\
&&\hspace*{-1mm} B^{(1)}_{x}=-2V_pB_0U^{(1)}_{sx}/V^{2}_{As},\label{17}\\
&&\hspace*{-1mm} B^{(1)}_{y}=-2V_pB_0U^{(1)}_{sy}/V^{2}_{As},\label{18}\
%&&\hspace*{-1mm}V_{p}=V_{As}/{\sqrt{2}}=V_A, \label{19}\
\end{eqnarray}

We now put $U^{(1)}_{sx}$ into the expression for $ B^{(1)}_{x}$ or $U^{(1)}_{sy}$ into the expression for $ B^{(1)}_{y}$, and obtain the linear dispersion relation:
\begin{eqnarray}
&&\hspace*{-1mm}V_{p}=V_{As}/{\sqrt{2}}=V_A, \label{19}\
\end{eqnarray}
where $V_A = {B_0}/{\sqrt{4\pi(m_eN_{e0}+m_pN_{p0})}}$ represents the effective EPSA speed.
Equation (\ref{19}) represents the general linear dispersion relation which represents the ultra-high frequency EPSA waves, where magnetic pressure $P_B=B_0^{2}/4\pi$
gives rise to the restoring force, and the net EP mass density
$\rho_T={m_eN_{e0}+m_pN_{p0}}$ provides the inertia.
\begin{figure}[!t]
\centering
\includegraphics[width=8cm]{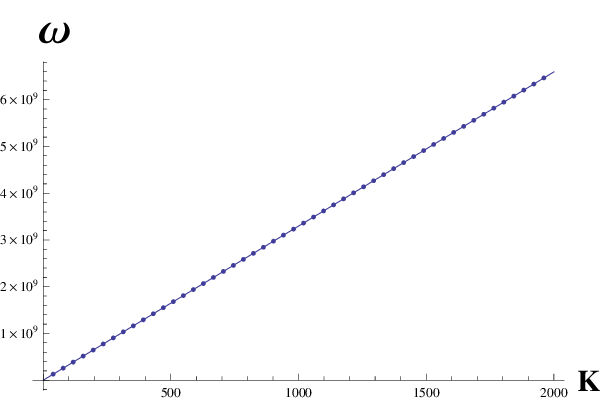}
\caption{Showing the variation of the wave frequency $\omega$ versus wavenumber $K$} \label{Fig1}
\end{figure}

\begin{figure}[!t]
\centering
\includegraphics[width=8cm]{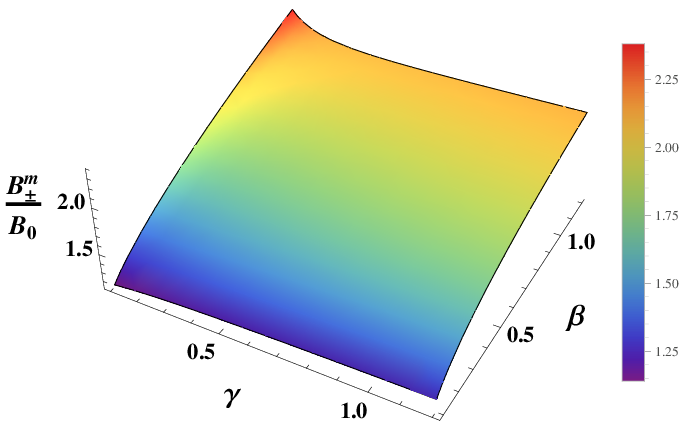}
\caption{Showing the variation of the maximum normalized amplitude ${{B}_\pm^m}/{B_0}$ of EPSA soliton with $\gamma$ and $\beta$} \label{Fig2}
\end{figure}

\begin{figure}[!t]
\centering
\includegraphics[width=8cm]{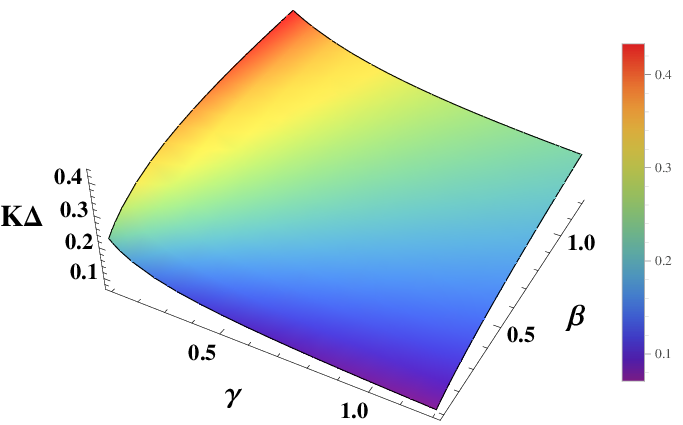}
\caption{Showing the variation of the normalized width $K\Delta$ of EPSA soliton with $\gamma$ and $\beta$} \label{Fig3}
\end{figure}

Now, substituting equation (\ref{13}) into equations
(\ref{7}) and (\ref{10}), and equating the coefficients of
$\epsilon^{2}$,  we obtain
\begin{eqnarray}
&&\hspace*{-10mm}\hspace*{15mm}U_{sz}^{(1)}=V_A|{B}_\pm|^2/2B^{2}_0,\label{20}\\
&&\hspace*{-10mm}\hspace*{15mm}N_{s}^{(1)}=N_{s0}|{B}_\pm|^2/2B^{2}_0,\label{21}
\end{eqnarray}
where ${B}_\pm=B_y^{(1)}\pm iB_x^{(1)}$ and $-$($+$) sign
represents the left-hand (right-hand) circularly polarized EPSA
waves. Again, as before, substitution of equation (\ref{13}) in equations (\ref{8}) and (\ref{9}) and (\ref{11}) and (\ref{12}), and equating the coefficients of $\epsilon^{5/2}$, we get another set of equations. Now, one can eliminate
$B_x^{(2)}$, $B_y^{(2)}$, $U_{sx}^{(2)}$,
$U_{sy}^{(2)}$ and obtain
\begin{eqnarray}
&&\hspace*{0mm}\hspace*{-8mm}2\partial_\tau B_x^{(1)}+\frac{V^{2}_A}{\omega_{cs}}\partial_\xi^2B_x^{(1)}+\frac{V_A }{2B_0^2}
\partial_\xi
[|{B}_\pm|^2B_y^{(1)}]=0,\label{22}\\
&&\hspace*{0mm}\hspace*{-8mm}2\partial_\tau B_y^{(1)}-\frac{V^{2}_A}{\omega_{cs}}\partial_\xi^2B_y^{(1)}+\frac{V_A }{2B_0^2}
\partial_\xi
[|{B}_\pm|^2B_x^{(1)}]=0,\label{23}
\end{eqnarray}
Multiplying equation (\ref{23}) by $\pm i$ then adding with
equation (\ref{22}) allows us to write
\begin{eqnarray}
&&\hspace*{-7mm}i\partial_\tau{B}_\pm\pm\frac{
V_A^2}{2\omega_{cs}}
\partial_\xi^2{B}_\pm+i\frac{V_A \
}{4B_0^2}\partial_\xi[|{B}_\pm|^2{B}_\pm]=0,
\label{24}
\end{eqnarray}
where $\partial_\xi[|{B}_\pm|^2{B}_\pm]$ is called
derivative nonlinear term. Equation (\ref{24}) is the well-known
derivative nonlinear Schr\"odinger equation, which
describes the nonlinear propagation of the high-frequency EPSA
wave in a magnetized electron-positron plasma. It is possible to
get above DNSE when higher-order physical variables are included
in the dynamics of wave packets as well as for exact parallel
propagation of right or left-handed polarized Alfv\'{e}n waves.
For obliquely propagating waves, one may derive a general form of
cubic nonlinear Schr\"{o}dinger equation \cite{Rajib2015}.

To solve the DNSE in (\ref{24}), we first normalize it by
transforming the dependent and independent variables as follows
\begin{eqnarray}
&&\hspace*{-0.01cm}\left.
\begin{array}{l}
{B}_\pm=2B_0\Psi_\pm,\\
\xi=\frac{V_A}{2\omega_{cs}}\zeta,\\
T=\frac{\tau }{2\omega_{cs}} .
\end{array}
\right\} \label{25}
\end{eqnarray}

These transformations of the dependent and independent variables
allow us to rewrite the DNSE (\ref{24}) in the form
\begin{eqnarray}
&&\hspace*{-0.01cm}i\partial_T
\Psi_\pm\pm\partial_\zeta^2\Psi_\pm+i\partial_\zeta[|\Psi_\pm|^2\Psi_\pm]=0,\label{26}
\end{eqnarray}\label{A}
where ${\partial}_T={\partial}/{\partial}T$ and
${\partial}_\zeta={\partial}/{\partial\zeta}$ and $-(+)$ sign in
front of the 2nd term of (\ref{26}) is for left (right) handed circularly
polarized EPSA waves.

To obtain the stationary solitonic solution of the DNSE
(\ref{26}), we consider a traveling co-ordinate
$X=\zeta{\pm}V_0T$, where $V_0$ is the wave phase speed normalized
by $V_A$ and $X$ is normalized by $(\frac{
V_A}{2\omega_{cs}})$. We, now apply the appropriate
boundary conditions, viz. $\Psi_\pm (X) \rightarrow 0$,
$d\Psi_\pm(X) /dX\rightarrow 0$, and $d^2\Psi_\pm
(X)/dX^2\rightarrow 0$ at $X\rightarrow \pm\infty$, and
use the basic property of complex variable $\Psi_-
\Psi_-^\ast=\Psi_+ \Psi_+^\ast=|\Psi_\pm|^2$.

The magnitude of the wave magnetic field ${B}_\pm$
(dimensional, where $|\Psi_\pm|$ is the normalized quantity), the
amplitude ${B}_\pm^m$ (dimensional, where $\Psi_\pm(X=0)$
is the normalized quantity) and the width $\Delta$ (in
dimensional form) are given by \cite{Ichikawa1979,Horton1996}
\begin{eqnarray}
&&\hspace*{7mm}\hspace*{-8mm}{B}_\pm=2B_0\sqrt{\left[
\frac{(1+\Gamma)\sqrt{2\beta}}{(\sqrt{1+2\Gamma})\cosh(X/\Delta)+
\sqrt{\Gamma}}\right]},\label{27}\\
&&\hspace*{7mm}\hspace*{-8mm}{B}_\pm^m=2B_0\sqrt{\left[
\frac{(1+\Gamma)\sqrt{2\beta}}{\sqrt{1+2\Gamma}+
\sqrt{\Gamma}}\right]},\label{28}\\
&&\hspace*{7mm}\hspace*{-8mm}\Delta=\frac{1}{2K}\sqrt{\left(\frac{\Gamma}{1+\Gamma}
\right)}, \label{29}
\end{eqnarray}
where $\Gamma=\beta/4\gamma$,
$\beta=\Omega/\omega_{cs}$, $\gamma=V_0/V_{A}$, and $X$ is the
dimensional space variable. We note that $K=2\pi/\lambda$ and
$\Omega=KV_0$ are used in (\ref{27})-(\ref{29}), where
$\lambda$ ($\Omega$) is the angular frequency dependent
wavelength of the EPSA waves.

\begin{figure}[!t]
\centering
\includegraphics[width=8cm]{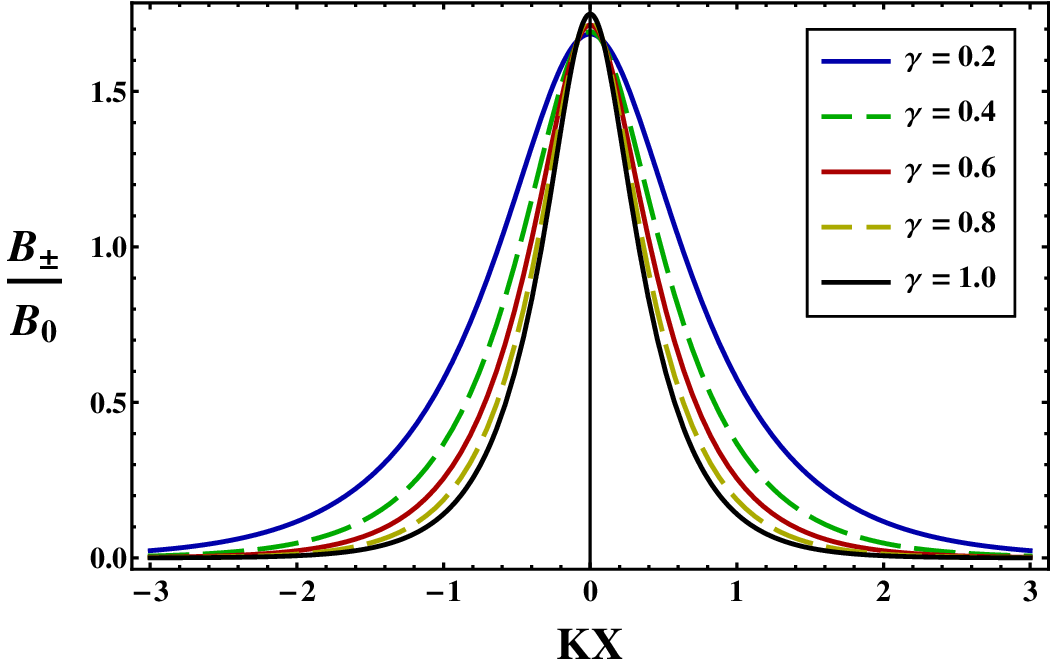}

\caption{Showing the solitary profile of EPSA waves along with the variation of $\gamma$  with $\beta=0.5$} \label{Fig4}
\end{figure}

\begin{figure}[!t]
\centering
\includegraphics[width=8cm]{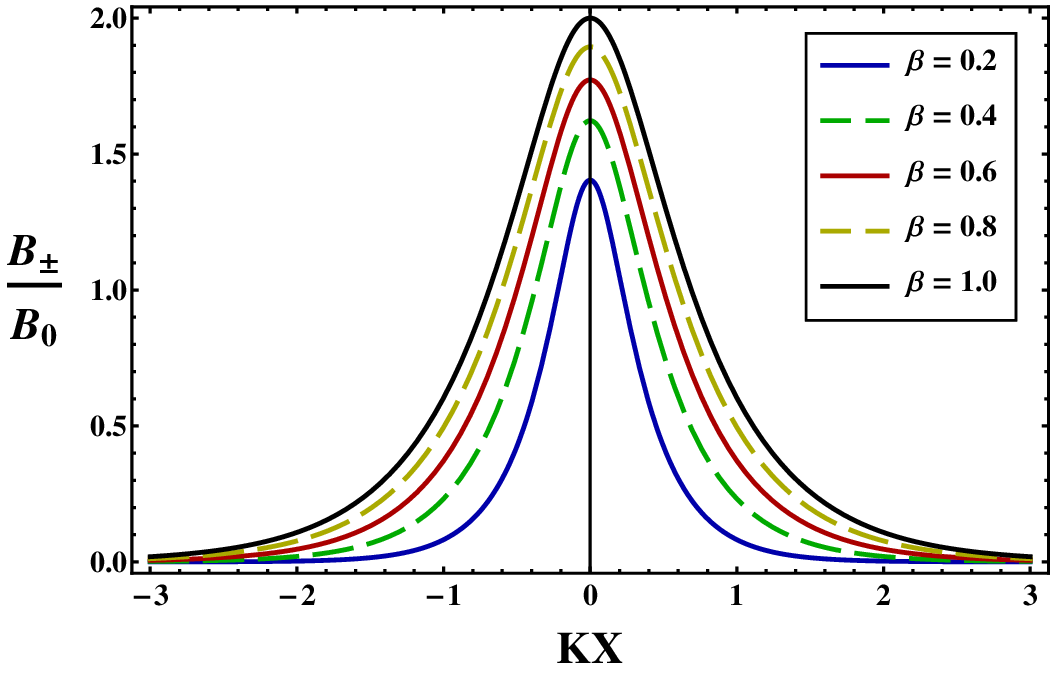}
\caption{Showing the solitary profile of EPSA waves along with the variation of $\beta$  with $\gamma=0.5$} \label{Fig5}
\end{figure}

\begin{figure}[!t]
\centering
\includegraphics[width=8cm]{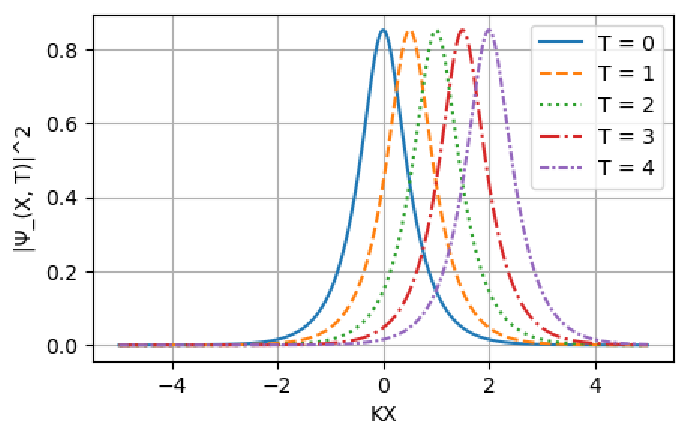}

\large{ (a)}
\vspace{0.0cm}

\includegraphics[width=8cm]{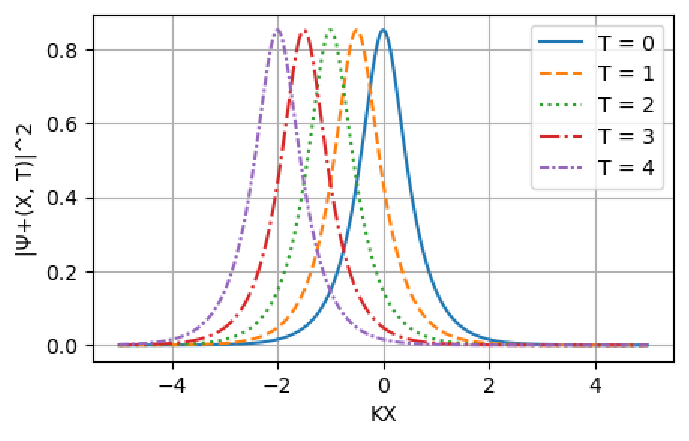}

\large{ (b)}
\caption{Time evolution of the (a) left-handed, and (b) right-handed circularly polarized EPSA solitons with $\beta=0.5$, $\gamma=0.5$, $V_0=0.5$, $K=1$, spatial grid spacing=0.0001, and time-step=0.01} \label{Fig6}
\end{figure}

\begin{figure}[!t]
\centering
\includegraphics[width=8cm]{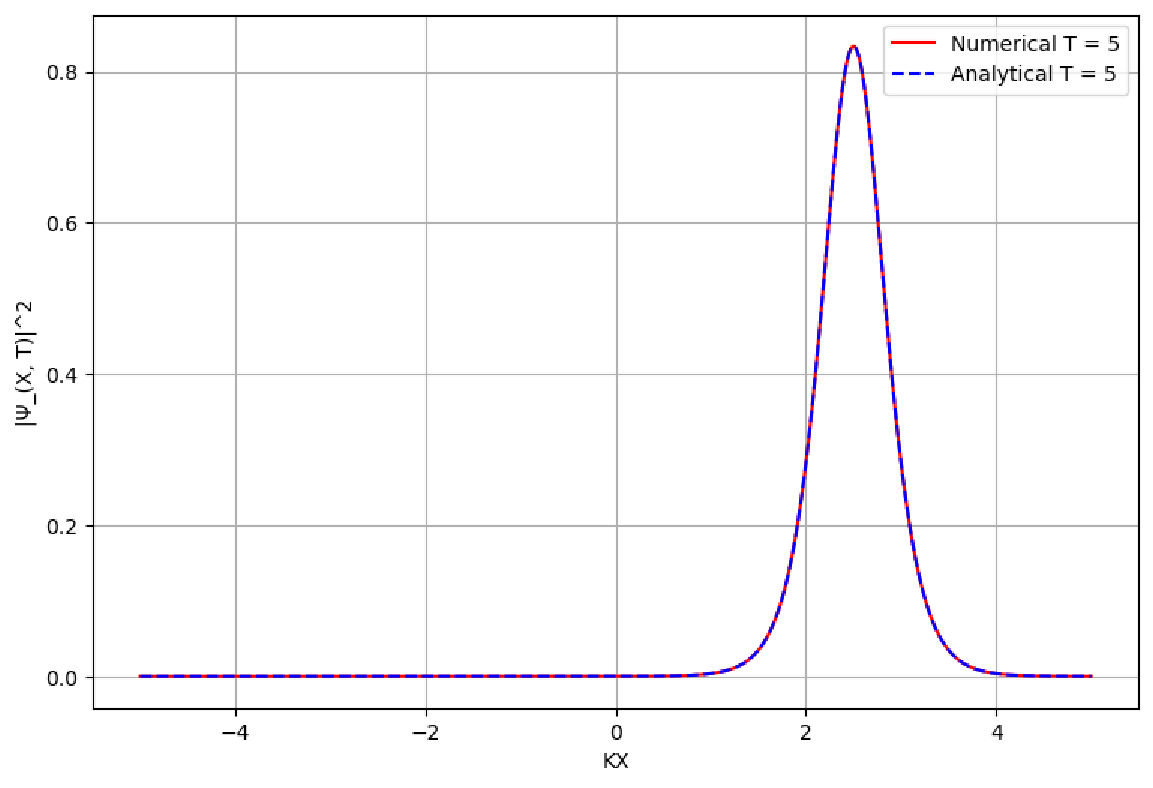}

\large{ (a)}
\vspace{0.0cm}

\includegraphics[width=8cm]{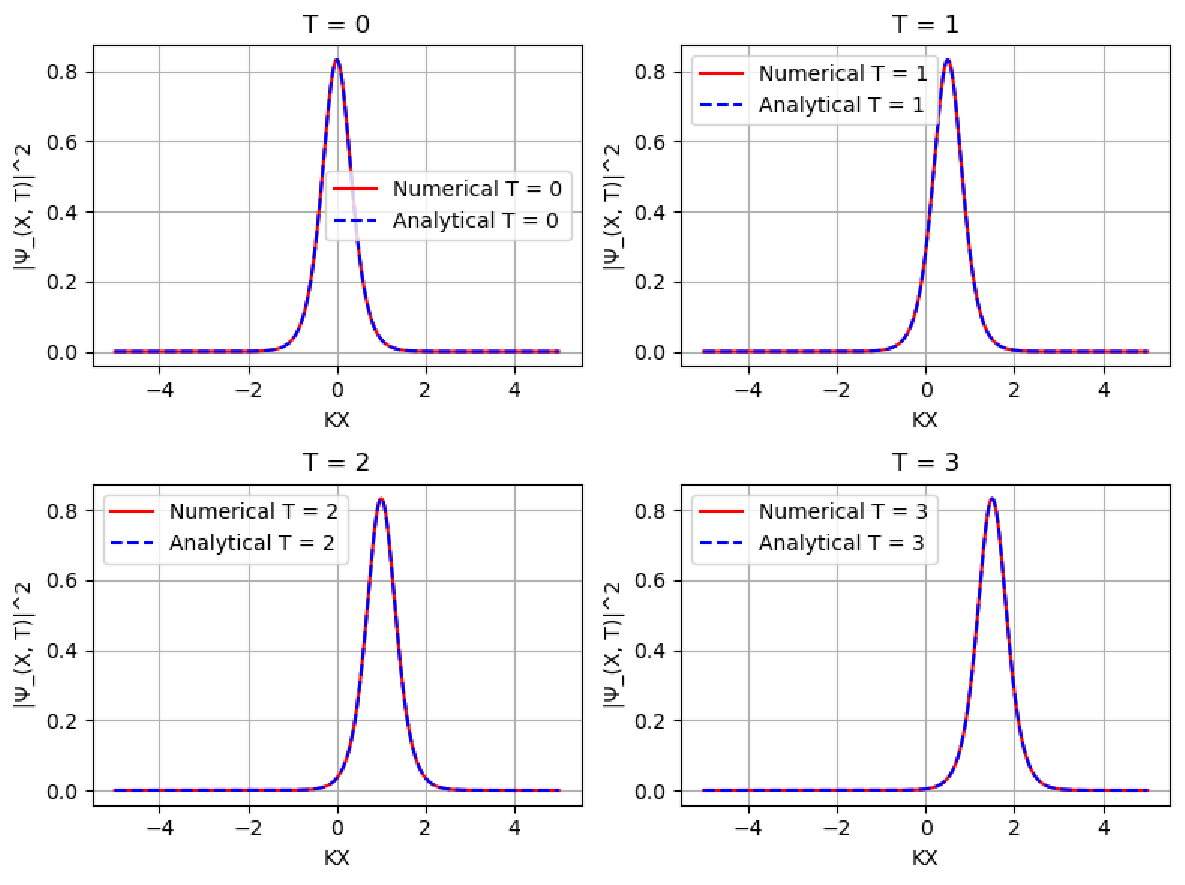}

\large{ (b)}
\caption{Comparison of analytical and numerical solutions of DNSE (\ref{26}) at (a) T=5 and (b) T=0,1,2,3 with  $\beta=0.5$, $\gamma=0.5$, $V_0=0.5$, $K=1$, spatial grid spacing=0.0001, and time-step=0.01 for left-handed circulary polarized EPSA waves} \label{Fig7}
\end{figure}

\section{Results and Discussion}
\label{NA}
We consider the propagation of electromagnetic perturbations in a pair EP plasma medium of opposite polarity of the same mass. We have analyzed the dispersion relation, maximum height, width, and solitary profile as well as compared it with the numerically solved solution in EPSA waves. For the numerical appreciation of the exact equation (\ref{27})-(\ref{29}), we consider a crab pulsar medium with EP plasma existing for external magnetic field $B_0$, mass density (approx.) $\rho_T$, and corresponding number density $N_{s0}=\rho_T/2m_s$ \cite{Haider2012}
\begin{table}[ht]
\caption{$B_0$, $\rho_T$, and $N_{s0}$ IN CRAB PULSAR MEDIUM\\\\} % title of Table
\centering      % used for centering table
\begin{tabular}{c c c c c}  % centered columns (4 columns)
\hline\hline                        %inserts double horizontal lines
Parameters && $~~Numerical~~Value$ & ~~~~~$Unit$  & ~~~~~$Ref.$\\ [0.1ex] % inserts table
 %heading
\hline\\                    % inserts single horizontal line
$B_0$ && $3.7\times10^{12}$ & ~~~~~$G$ & ~~~~~\cite{Bhatt1985}  \\\\   % inserting the body of the table
$\rho_T$ && $10^{12}-10^{14}$ & ~~~~~$g/cm^3$ & ~~~~~\cite{Haider2012}  \\\\
$N_{s0}$ && $5.5\times(10^{37}-10^{39})$ & ~~~~~$cm^{-1}$ & ~~~~~\cite{Haider2012}  \\ [1ex]
%3 & 31 & 25  & 415  \\
%4 & 35 & 144 & 2356 \\
%5 & 45 & 300 & 556 \\       % [1ex] adds vertical space
\hline     %inserts single line
\end{tabular}
\label{T1}  % is used to refer this table in the text \end{table}
\end{table}

\begin{table}[!ht]
\caption{NORMALIZED MAXIMUM HEIGHT (APPROXIMATE), ${{B}_\pm^m}/{B_0}$ AND WIDTH, $K\Delta$ for DIFFERENT VALUES OF $\beta$ WHEN $\gamma=0.5$\\} % title of Table
\centering      % used for centering table
\begin{tabular}{c c c c c}  % centered columns (4 columns)
\hline\hline                        %inserts double horizontal lines
Parameters ~~~~~&& $\beta=0.2$ & ~~~~~~~$\beta=0.6$ & ~~~~~$\beta=1.0$ \\ [0.1ex] % inserts table
 %heading
\hline\\                    % inserts single horizontal line
${{B}_\pm^m}/{B_0}$ && 1.40 & ~~~~~~~1.77 & ~~~~~~~2.0  \\\\   % inserting body of the table
${K\Delta}$ && 0.15 & ~~~~~~~0.24 & ~~~~~~~0.29  \\ [1ex]
%3 & 31 & 25  & 415  \\
%4 & 35 & 144 & 2356 \\
%5 & 45 & 300 & 556 \\       % [1ex] adds vertical space
\hline     %inserts single line
\end{tabular}
\label{T2}  % is used to refer this table in the text \end{table}
\end{table}

\begin{itemize}
\item {\bf{\textit{Linear dispersion relation}}}:  It is clearly seen that that wave angular frequency $\omega$ is increased linearly with wave number $K$ as expected in Fig. \ref{Fig1}. The observed radio emission (typically radio emission range $\omega=10^9-10^{11} Hz)$
from pulsars, which are magnetized neutron stars, is generated in an EP pair plasma and must propagate
through such plasma as it escapes \cite{Manchester1977}.

\begin{table}[t]
\caption{NORMALIZED MAXIMUM HEIGHT (APPROXIMATE), ${{B}_\pm^m}/{B_0}$ AND WIDTH, $K\Delta$ for DIFFERENT VALUES OF $\gamma$ WHEN $\beta=0.5$\\} % title of Table
\centering      % used for centering table
\begin{tabular}{c c c c c}  % centered columns (4 columns)
\hline\hline                        %inserts double horizontal lines
Parameters ~~~~~&& $\gamma=0.2$ & ~~~~~~~$\gamma=0.6$ & ~~~~~~~$\gamma=1.0$ \\ [0.1ex] % inserts table
 %heading
\hline\\                    % inserts single horizontal line
${{B}_\pm^m}/{B_0}$ && 1.67 & ~~~~~~~1.70 & ~~~~~~~1.74  \\\\   % inserting body of the table
${K\Delta}$ && 0.31 & ~~~~~~~0.21 & ~~~~~~~0.17  \\ [1ex]
%3 & 31 & 25  & 415  \\
%4 & 35 & 144 & 2356 \\
%5 & 45 & 300 & 556 \\       % [1ex] adds vertical space
\hline     %inserts single line
\end{tabular}
\label{T3}  % is used to refer this table in the text \end{table}
\end{table}

\item{\bf{\textit{Variation of ${{B}_\pm^m}/{B_0}$ against $\gamma$ and $\beta$}}}:  The maximum potential $2\Psi_{\pm}={{B}_\pm^m}/{B_0}$ increases very slightly with the increase of $\gamma$ and gradual increase has been seen for the variation of $\beta$. It shows the maximum value for the combination of low $\gamma$ and higher $\beta$ values (see Fig. \ref{Fig2} and Tables \ref{T2} $\&$ \ref{T3} ). We may conclude that the variation of $\gamma$ is insignificant here.

\item {\bf{\textit{Variation of ${K\Delta}$ against $\gamma$ and $\beta$}}}: The width $K\Delta$, with $\gamma$, gradually decreases until  $\gamma= 0.6$, and after that, it remains almost constant (see Fig. \ref{Fig3}). In stark contrast, however, $K\Delta$, gradually increases with an
increase in $\beta$ shown in Fig. \ref{Fig3} and Tables \ref{T2} $\&$ \ref{T3} ).

\item {\bf{\textit{Variation of the solitonic profile, ${{B}_\pm}/{B_0}$ versus $KX$ for different values of $\gamma$ and $\beta$}}}:  The solitonic profile is formed due to the balance between the nonlinear coefficient and dispersion coefficient. In the considered plasma system, solitary
waves have been formed with a narrow width and almost constant height for the high value of $\gamma$ (see Fig. \ref{Fig4}), and it is seen that both the
thickness and height gradually increase with the increase in $\beta$ as shown in Fig. \ref{Fig5}.

\item{\bf{\textit{Time evolution of solitary waves}}}: By modifying equation (\ref{27}), $\Psi_{\pm}(X\ne 0, T)={B}_\pm/2B_0=\sqrt{\left[
\frac{(1+\Gamma)\sqrt{2\beta}}{(\sqrt{1+2\Gamma})\cosh[(X{\mp}V_0T)/\Delta]+\sqrt{\Gamma}}\right]}exp(iKX)$, we can get the time evolution of the solitary waves. At $X=0$ and $T=0$ with $K=1$ (insignificant here at X=0 as $exp(iKX)=1$), we get the stationary solution the same as (\ref{27}). For left (right) handed circularly polarized shown EPSA in Fig. \ref{Fig6}-a(\ref{Fig6}-b), the shape of the solitary waves is fully conserved. All are considered in normalized cases.

\item{\bf{\textit{Comparison of analytical and numerical solutions}}}: The finite difference method (FDM) has been successfully applied to finding the numerical solution of DNSE  (\ref{26}) \cite{C2007,M2010}. In Fig. \ref{Fig7}, one soliton solutions were obtained at different times. The results showed that FDM and analytical solitons solutions behaved similarly. It is noted that for mathematical convenience physical relevance (secant hyperbolic) and numerical stability, we neglected the $\sqrt{\Gamma}$ in the denominator of the solution $\Psi_{\pm}(X\ne 0, T)$.

\end{itemize}

\section{Conclusion}
\label{D}
We have considered an EP system collisionless magnetized EP plasma to study the nonlinear propagation of solitary
waves for which the mass ratio of positron-to-electron $\alpha$ is equal to one. The following are the investigation's predicted findings:

\begin{itemize}
\item{The EP medium under consideration
supports extremely high phase speed, high-frequency EPSA waves (propagating parallel
to the external magnetic field $B_0$), in which the magnetic pressure
gives rise to the restoring force, and the net EP mass density
provides the inertia.}

\item{ From the linear dispersion relation in equation  (\ref{19}), it is concluded that crab pulsar emits radio frequency as expected depicted in Figure \ref{Fig1}}.

\item{The EP plasma medium under consideration supports a shear  Alfv\'enic wave of solitary profile that is associated
with exactly propagating with the external magnetic field.}

\item{Only Hump shaped solitary profiles are observed. The height (width) of the EPSA hump shape solitary waves maintain almost constant (decreases) with the increase of $\gamma$ with a fixed value of $\beta=0.5$ (Figure \ref{Fig4}). In stark contrast, however, in solitary profile, as we increase $\beta$, both their height and width increase with a fixed value of $\gamma=0.5$} (Figure \ref{Fig5}).
    
\item{The time evolution of solitary waves indicates that solitons are formed and maintain their shape with time (Figure \ref{Fig6}). In addition, analytical and numerical solutions have a good agreement with each other (Figure \ref{Fig7}).}
\end{itemize}

%We note that the nonlinear analysis presented here is not valid when $\theta=90^{\circ}$ (perpendicular propagation), in which case one should
%derive the Korteweg–de Vries (KdV) equation, and examine
%the properties of the compressional Alfv\'{e}n solitons \cite{Rajib2017}. But, except for the perpendicular propagation $\theta=90^{\circ})$ \cite{Rajib2018}, our theory is valid for arbitrary
%values of $\beta$ and $\gamma$. This means that our present work is not only valid for
%the opposite polarity EP plasma medium, but also valid for any kind of
%two-component plasma systems, particularly dust
%($\alpha=Z_nm_p/Z_pm_n$) and electron-ion ($\alpha\le 1/1836$) plasmas.

It is important to note that the nonlinear analysis presented in this study is not applicable when the propagation angle is perpendicular (i.e., $\theta=90^{\circ}$), in which case the derivation of the KdV equation is necessary, and the properties of compressional Alfvén solitons should be examined \cite{Rajib2022a,Rajib2017}. However, except for perpendicular propagation (i.e., parallel propagation when $\theta=0^{\circ}$) \cite{Rajib2018}, our theory remains valid for any arbitrary values of $\beta$ and $\gamma$. This implies that our current work is not only applicable to opposite polarity EP plasma media but also extends its validity to any two-component plasma systems, including dust ($\alpha=Z_nm_p/Z_pm_n$) and electron-ion ($\alpha\le 1/1836$) plasmas.

%By way of conclusion, we may say that our theoretical investigation should be helpful in a better understanding of the characteristics of small but finite amplitude electromagnetic disturbances that are %ubiquitous in a laboratory as well as space plasmas, where opposite polarity plasma components are available such as the solar wind \cite{Clem2000,Adriani2009,Adriani2011}, the magnetosphere of Earth %\cite{Ackermann2012,Aguilar2014}, pulsars magnetosphere \cite{Profumo2012}, and microquasars \cite{Siegert2016}.

In conclusion, our theoretical investigation contributes to an enhanced understanding of the features of small yet finite amplitude electromagnetic disturbances prevalent in laboratory and space plasmas. This is particularly relevant in scenarios where opposite polarity plasma components are present, such as in the solar wind \cite{Clem2000,Adriani2009,Adriani2011}, Earth's magnetosphere \cite{Ackermann2012,Aguilar2014}, pulsars' magnetospheres \cite{Profumo2012}, and microquasars \cite{Siegert2016}.\\

\section*{ACKNOWLEDGMENTS}
I express my gratitude to Prof A A Mamun for his unwavering support and for affording me the autonomy to function as an independent researcher.
%This research article is dedicated to the little princess, \textbf{\textit{Reeha Tanvir's first birthday is on December 26, 2021}}, to keep the day memorable. One of the authors, T. I. Rajib is grateful to Jahangirnagar %University (JU) for the financial support through the JU Research Grant (2021-2022). This fund has given him %the freedom to focus on his research. Moreover, we acknowledge Prof A A Mamun for his continuous %support.
\section*{DATA AVAILABILITY}
Data sharing does not apply to this article as no new data were created or analyzed in this study.

%\section*{References}

%\end{sloppypar}
\end{document}